\documentclass[preprint,5p,times,twocolumn,compress]{elsarticle}
\usepackage[intlimits]{amsmath}
\usepackage{amssymb}
\usepackage[T1]{fontenc}
\usepackage[normalem]{ulem} 
\usepackage{color}
\usepackage{mathtools} 
\usepackage{pbox}
\usepackage{enumitem}
\makeatletter
\def\ps@pprintTitle{%
 \let\@oddhead\@empty
 \let\@evenhead\@empty
 \def\@oddfoot{}%
 \let\@evenfoot\@oddfoot}
\makeatother

\usepackage{lipsum}
\usepackage[hidelinks]{hyperref}

\begin{document}

%%fakesection frontmatter
\begin{frontmatter}



  \title{On precision holography for the circular Wilson loop in $AdS_{5}\times S^{5}$}
\author{Li Botao}
\author{Daniel Medina-Rincon}
\address{
\textit{Institut f\"ur Theoretische Physik,\\
Eidgen\"ossische Technische Hochschule Z\"urich,\\
Wolfgang-Pauli-Strasse 27, 8093 Z\"urich, Switzerland}\\
\texttt{\upshape libota@student.ethz.ch\quad meddanie@phys.ethz.ch}
}

%%%%%%%%%%%%%%%%%%%%%%%%%%%%%%%%%%%%%%%%%%%%%%%%%%%%%%%%%%%%%%%%%%%%
%%%%%%%%%%%%%%%%%%%%%%%%%%%%%%%%%%%%%%%%%%%%%%%%%%%%%%%%%%%%%%%%%%%%
%%%%%%%%%%%%%%%%%%%%%%%%%%%%%%%%%%%%%%%%%%%%%%%%%%%%%%%%%%%%%%%%%%%%

\begin{abstract}
The string theory calculation of the $\frac{1}{2}$-BPS circular Wilson loop of $\mathcal{N}=4$ SYM in the planar limit at next to leading order at strong coupling is revisited in the ratio of its semiclassical string partition function and the one dual to a latitude Wilson loop with trivial expectation value. After applying a conformal transformation from the disk to the cylinder, this problem can be approached by means of the Gel'fand-Yaglom formalism. Using results from the literature and the exclusion of zero modes from a modified Gel'fand-Yaglom formula, we obtain matching with the known field theory result. As seen in the phaseshift method computation, non-zero mode contributions cancel and the end result comes from the zero mode degeneracies of the latitude Wilson loop. \end{abstract}

\end{frontmatter}

%%%%%%%%%%%%%%%%%%%%%%%%%%%%%%%%%%%%%%%%%%%%%%%%%%%%%%%%%%%%%%%%%%%%
%%%%%%%%%%%%%%%%%%%%%%%%%%%%%%%%%%%%%%%%%%%%%%%%%%%%%%%%%%%%%%%%%%%%
%%%%%%%%%%%%%%%%%%%%%%%%%%%%%%%%%%%%%%%%%%%%%%%%%%%%%%%%%%%%%%%%%%%%

\section{Introduction}
Wilson loops have played an important role in the study of the correspondence between $\mathcal{N}=4$ SYM field theory with gauge group $SU(N)$ and type IIB strings in $AdS_{5}\times S^{5}$ \cite{Maldacena:1997re}, as there exists a known description at both sides of the duality \cite{Maldacena:1998im,Rey:1998ik}. The 1/2-BPS circular Wilson is of particular relevance in this correspondence as its expectation value in the planar limit has been calculated in field theory to all orders in the 't Hooft coupling by means of diagram resummation \cite{Erickson:2000af} and supersymmetric localization \cite{Pestun:2007rz}. Exact results at all orders have allowed for precision tests of the duality and for the development of holographic techniques to compute these quantities.

In the string theory side, the expectation value of the Wilson loop is given by the partition function of a string whose worldsheet encloses the Wilson loop contour at the boundary of $AdS$ \cite{Polyakov:1997tj}. Evaluation of the string partition function can be performed using the semiclassical approximation. In this case the leading contribution corresponds to the regularized area of the worldsheet describing the string classical solution \cite{Drukker:1999zq}, while the next to leading order is described by small fluctuations around the classical solution. Wilson loop string theory calculations have perfectly matched field theory predictions at leading order, however,  matching at next to leading order has been achieved only recently.

Based on the pioneering work in \cite{Drukker:2000ep}, early attempts at computing the 1-loop string partition function for the circular Wilson loop lead to mismatches with the field theory prediction \cite{Kruczenski:2008zk,Kristjansen:2012nz,Buchbinder:2014nia}. A major obstacle for this computation is the unknown normalization of the string partition function\footnote{Progress on this issue has been recently reported in \cite{Giombi:2020mhz}.}. To circumvent this problem, \cite{Forini:2015bgo,Faraggi:2016ekd} considered the ratio of the 1/2-BPS circular Wilson loop with a 1/4-BPS Wilson loop with latitude angle $\theta_0\in[0,\pi/2)$ in the Gel'fand-Yaglom  formalism and obtained the field theory prediction plus an additional term. This issue was resolved to first order for infinitesimal small angle $\theta_0$ \cite{Forini:2017whz}, and for all values of $\theta_0$ using the phaseshift method \cite{Cagnazzo:2017sny}, finally recovering the field theory prediction for this ratio. 

Building on these results, in \cite{Medina-Rincon:2018wjs} the 1/2-BPS Wilson loop was computed in string theory at 1-loop by considering its ratio with a ``special'' latitude Wilson loop with trivial expectation value \cite{Zarembo:2002an}, finally achieving a precise matching with field theory\footnote{Matching with field theory was also recently obtained in \cite{Giombi:2020mhz} introducing a normalization prefactor and using Heat kernel on an individual loop.}. Using the phaseshift method, the final answer was shown to come exclusively from the zero modes due to degeneracies in the classical string solution of the latitude. 

In this letter, we revisit the calculation of \cite{Medina-Rincon:2018wjs} using this time the Gel'fand-Yaglom formalism \cite{doi:10.1063/1.1703636}. This technique is widely used for the evaluation of determinants due to its relatively simple procedure, but has not been used for Wilson loop string computations in cases where zero mode exclusion is required. The calculation of functional determinants with a zero eigenvalue has been widely studied in the literature since the early work \cite{McKane:1995vp}, and has led to modified Gel'fand-Yaglom expressions coming from different approaches \cite{,Kleinert:1998rz,kleinert1999functional,falco2017functional,KIRSTEN2003502,Kirsten:2005di}. By revisiting this Wilson loop calculation using a different technique, we aim at enriching the toolbox for precision holography of Wilson loops.

This paper is organized as follows. In section \ref{Sec2} we describe the main features of the circular and ``special'' latitude Wilson loops in field theory and string theory. Section \ref{Sec3} presents the setup for the 1-loop string theory calculation of the ratio of Wilson loops following \cite{Medina-Rincon:2018wjs}, while section \ref{Sec4} concerns the evaluation of determinants using Gel'fand-Yaglom results and zero mode exclusion. Finally, section \ref{Sec5} presents our concluding remarks. \ref{GYAppendex} contains some results from the literature for Wilson loops in the Gel'fand-Yaglom method.

%%%%%%%%%%%%%%%%%%%%%%%%%%%%%%%%%%%%%%%%%%%%%%%%%%%%%%%%%%%%%%%%%%%%
%%%%%%%%%%%%%%%%%%%%%%%%%%%%%%%%%%%%%%%%%%%%%%%%%%%%%%%%%%%%%%%%%%%%
%%%%%%%%%%%%%%%%%%%%%%%%%%%%%%%%%%%%%%%%%%%%%%%%%%%%%%%%%%%%%%%%%%%%

\section{Circular and latitude Wilson loops in $\mathcal{N}=4\, \rm{SYM}$ and $AdS_{5}\times S^{5}$}\label{Sec2}
In $\mathcal{N}=4$ SYM with gauge group $SU(N)$ the Wilson loop is defined as \cite{Maldacena:1998im}
\begin{align}
W\left( {C;\bf{n}} \right) = \frac{1}{N}\left\langle {\text{tr\ P}\exp \left[ {i\int\limits_C {\left( {{{\dot x}^\mu }{A_\mu } + i\left| {\dot x} \right|{n^I}{\Phi _I}} \right)d\tau } } \right]} \right\rangle ,
\end{align}
where it is described by the contour $C$ parametrized as $x^{\mu}\left(\tau\right)$ and by the unit norm 6-vector $n^{I}$ in $S^5$ describing the coupling to scalars. 

The two Wilson loops considered here belong to a family of latitude Wilson loops parametrized by an angle $\theta_0$, where the 1/2-BPS circle corresponds to the configuration where $\theta_{0}=0$, while $\theta_0=\pi/2$ for the 1/4-BPS ``special'' latitude. Both Wilson loops have the unit circle for $C$, while the coupling to scalars is described by
\begin{align}\label{md1}
{\bf{n}}=\left(\bf{k}\cos\theta,\cos\varphi \sin\theta,\sin\varphi\sin\theta\right)
\end{align}
with the parameters
\begin{align}
&\text{C:}& \theta&=0, & {\bf{k}}&=\left(1,0,0,0\right),& \varphi&=\text{any},\\
&\text{L:}& \theta&=\pi/2, & {\bf{k}}&=\text{any}, & \varphi&=\tau,
\label{qft1}
\end{align}
for the circular and latitude Wilson loops, respectively. We see that for the circle, as $\theta=0$, $\bf{n}$ is at the north pole where $\varphi$ is arbitrary. Meanwhile, for the latitude, at $\theta=\pi/2$ the $S^{3}\subset S^{5}$ shrinks allowing the unit 4-vector $\bf{k}$ to take any value.

The expectation values for these Wilson loops in the planar limit are \cite{Erickson:2000af,Pestun:2007rz,Zarembo:2002an,Drukker:2000rr}
\begin{align}\label{qftpredictions}
W_{\text{C}}=\frac{2}{\sqrt{\lambda}}I_{1}\left(\sqrt{\lambda}\right)\overset{\lambda\to\infty}{\longrightarrow} \sqrt{\frac{2}{\pi}}\lambda^{-3/4}e^{\sqrt{\lambda}}, && W_{\text{L}}=1.
\end{align}

In the $AdS_{5}\times S^{5}$ background described by the metric
\begin{align}\label{Targetspacemetric}
d{s^2} = \frac{{d{Z^2} + dX_\mu ^2}}{{{Z^2}}} + d{\theta ^2} + {\sin ^2}\theta d{\varphi ^2} + {\cos ^2}\theta d\Omega _{{S^3}}^2,
\end{align}
the string configurations dual to these Wilson loops correspond to 
\begin{align}
\text{C},\text{L}:&& X^{\mu}=\left(\frac{\cos\tau}{\cosh\sigma},\frac{\sin\tau}{\cosh\sigma},0,0\right)&& Z=\tanh\sigma,
\end{align}
in $AdS_5$, while in $S^{5}$
\begin{align}\label{md2}
\text{C}: \theta=0, &&\qquad\qquad
\text{L}: \varphi=\tau, \quad\quad \cos\theta=\tanh\sigma.
\end{align}
It is easy to see that at the $AdS$ boundary, where $\sigma=0$, the strings end in the unit circle. Meanwhile, the 1/2-BPS circle remains only at a point in $S^{5}$ while the 1/4-BPS ``special'' latitude extends inside the sphere. Note that for the latitude, as $\sigma\to0$, the volume of $S^{3}$ shrinks to zero size and thus the angles in $S^{3}$ can take arbitrary values. The later is compatible with the field theory description \eqref{qft1} with arbitrary $\bf{k}$ and explicitly shows that the ``special'' latitude corresponds to a degenerate 3-parametric family of solutions. These latitude string solutions are different for $\sigma\neq0$ and correspond to non-trivial moduli \cite{Zarembo:2002an}.

%%%%%%%%%%%%%%%%%%%%%%%%%%%%%%%%%%%%%%%%%%%%%%%%%%%%%%%%%%%%%%%%%%%%
%%%%%%%%%%%%%%%%%%%%%%%%%%%%%%%%%%%%%%%%%%%%%%%%%%%%%%%%%%%%%%%%%%%%
%%%%%%%%%%%%%%%%%%%%%%%%%%%%%%%%%%%%%%%%%%%%%%%%%%%%%%%%%%%%%%%%%%%%

\section{The perturbative string theory computation}\label{Sec3}
\subsection{Setup}
In order to evaluate the string partition function it is necessary to gauge fix, this is done by choosing the worldsheet metric to be the metric induced by the classical solution \cite{Drukker:2000ep}. This results in the following worldsheet metrics 
\begin{align}\label{metric1}
ds_{\rm{ws}}^2 ={\Omega ^2}\left( {d{\tau ^2} + d{\sigma ^2}} \right)
\end{align}
with
\begin{align}\label{metric2}
{\Omega ^2_{\text{C}}} = \frac{1}{{{{\sinh }^2}\sigma }}, && {\Omega ^2_{\text{L}}} = \frac{1}{{{{\sinh }^2}\sigma }} + \frac{1}{{{{\cosh }^2}{\sigma}}},
\end{align}
for the circle and latitude, respectively.

Following the proposal of \cite{Forini:2015bgo,Faraggi:2016ekd}, we consider the ratio of these Wilson loops and omit contributions from ghosts and longitudinal modes as the configurations considered have the same topology. Note that this assumption for ratios of circular Wilson loops is also supported by the results of \cite{Forini:2017whz,Cagnazzo:2017sny,Medina-Rincon:2018wjs,Medina-Rincon:2019bcc,David:2019lhr}.

Semiclassical evaluation of the string partition function for the circle results in
\begin{align}
Z_{\text{C}}=\int\mathcal{D}\Phi\ e^{-\frac{\sqrt{\lambda}}{2\pi}S[\Phi]}\approx e^{-\frac{\sqrt{\lambda}}{2\pi}S\left[\Phi_{\text{cl}}\right]}\ \rm{Sdet}^{-1/2} \mathbb{K}_{\text{C}},
\end{align}
where $\Phi$ denotes the different fields, $\Phi_{\text{cl}}$ the classical solution, while $\rm{Sdet}\,\mathbb{K}_{\text{C}}$ corresponds to bosonic and fermionic determinants from second order fluctuations. The equivalent expression for the latitude is more elaborate as zero modes from the moduli in the classical solution are traded by collective coordinates in the path integral \cite{Medina-Rincon:2018wjs}
\begin{align}
Z_{\text{L}}\approx e^{-\frac{\sqrt{\lambda}}{2\pi}S\left[\Phi_{\text{cl}}\right]} \rm{Sdet'}^{-1/2} \mathbb{K}_{\text{L}}\int\prod_{n=1}^{3}\frac{\lambda^{1/4}}{2\pi}d\varphi_{n}\, {\det_{ij}}^{1/2}\left<\frac{\partial\Phi_{\text{cl}}}{\partial\varphi_{i}},\frac{\partial\Phi_{\text{cl}}}{\partial\varphi_{j}} \right>,
\end{align}
where there is additional integration over the moduli space of solutions each with a factor\footnote{This factor is chosen such that the path integral over fluctuations with non-zero eigenvalues leads to the determinant of the differential operator without any additional multiplicative factors coming from the string action. For more details see \cite{Medina-Rincon:2018wjs}.}  of $\lambda^{1/4}/2\pi$, a Jacobian from the change of variables\footnote{In fact, $\partial\Phi_{\text{cl}}/\partial\varphi_i$ correspond to zero modes of $\mathbb{K}=\delta^{2}S/\delta\Phi^{2}|_{\text{cl}}$. To see this differentiate the equations of motion $0=\frac{\partial}{\partial\varphi_i}\frac{\delta S}{\delta \Phi}\big{|}_{\text{cl}}=\frac{\delta^2 S}{\delta\Phi\delta\Phi}\frac{\partial \Phi}{\partial\varphi_i}\big{|}_{\text{cl}}=\mathbb{K}\frac{\partial \Phi_{\text{cl}}}{\partial\varphi_i}$.} and $\text{Sdet}'$ represents the corresponding fluctuation determinants with the zero modes excluded. 

It is easy to see that the string action at the classical solution satisfies
\begin{align}
\text{C}:\quad S\left[\Phi_{\text{cl}}\right]=-2\pi,&& \text{L}:\quad S\left[\Phi_{\text{cl}}\right]=0, 
\end{align}
which perfectly reproduce the exponential behaviours in \eqref{qftpredictions}. We will now focus on the different contributions to the ratio at next to leading order for large $\lambda$. 

%%%%%%%%%%%%%%%%%%%%%%%%%%%%%%%%%%%%%%%%%%%%%%%%%%%%%%%%%%%%%%%%%%%%

\subsection{Integration over moduli}\label{moduli}

Integration over the angles of $S^{3}$ corresponding to the moduli of the latitude classical solution follows from \eqref{md1} and \eqref{md2}
\begin{align}
\frac{\partial\bf{n}}{\partial\varphi_i}\bigg{|}_{\text{cl}}=\left(\frac{\partial\bf{k}}{\partial\varphi_i}\tanh\sigma,0,0\right),
\end{align}
resulting in
\begin{align}\label{esa1}
\int_{S^{3}}\prod_{n=1}^{3}d\varphi_{n}\, {\det_{ij}}^{1/2}\left<\frac{\partial\bf{n}}{\partial\varphi_{i}},\frac{\partial\bf{n}}{\partial\varphi_{j}} \right>=2\pi^{2}\left<\psi_0|\Omega_{\text{L}}^{2}|\psi_0\right>^{3/2}.
\end{align}
In the above $\psi_0=\tanh\sigma$ and we used the definition of the inner product implied by the worldsheet metric \eqref{metric1}
\begin{align}
\left<\psi_1\big{|}\Omega^{2}\big{|}\psi_2\right>=\int_{0}^{2\pi}\int_{0}^{\infty}\Omega^{2}\psi_1 \psi_2 \, d\sigma d\tau.
\end{align}
Explicit evaluation shows that the right hand side of \eqref{esa1} is finite, consistent with the normalizability of the zero mode $\psi_0$.

From the above, we obtain for the ratio
\begin{align}
\frac{Z_{\text{C}}}{Z_{\text{L}}}=e^{\sqrt{\lambda}}\lambda^{-3/4}4\pi\frac{\rm{Sdet'}^{1/2}\mathbb{K}_{\text{L}}}{\rm{Sdet}^{1/2}\mathbb{K}_{\text{C}}}\left<\psi_0|\Omega_{\text{L}}^{2}|\psi_0\right>^{-3/2}.
\end{align}

%%%%%%%%%%%%%%%%%%%%%%%%%%%%%%%%%%%%%%%%%%%%%%%%%%%%%%%%%%%%%%%%%%%%

\subsection{Functional determinants}\label{GYSection}

Expansion of the Green-Schwarz action to second order in fluctuations around the classical solution leads to the following ratio of determinants for each partition function
\begin{align}\label{Sdet}
\rm{Sdet} \mathbb{K} = \frac{{{{\det }^3}{\mathcal{K}_1}{{\det }^3}{\mathcal{K}_2}\det {\mathcal{K}_{3 + }}\det {\mathcal{K}_{3 - }}}}{{{{\det }^4}{\mathcal{D}_ + }{{\det }^4}{\mathcal{D}_ - }}},
\end{align}
where the operators above are given in terms of \cite{Forini:2015bgo,Faraggi:2016ekd} 
\begin{align}
{{\widetilde{\mathcal K}}_1} &=  - \partial _\tau ^2 - \partial _\sigma ^2 + \frac{2}{{{{\sinh }^2}\sigma }},\\
{{\widetilde{\mathcal K}}_2} &=  - \partial _\tau ^2 - \partial _\sigma ^2 - \frac{2}{{{{\cosh }^2}\left( {\sigma  + {\sigma _0}} \right)}},\\
{{\widetilde{\mathcal K}}_{3 \pm }} &=  - \partial _\tau ^2 - \partial _\sigma ^2 \pm 2i\left( {\tanh \left( {2\sigma  + {\sigma _0}} \right) - 1} \right){\partial _\tau }\nonumber \\
&\quad+ \left( {\tanh \left( {2\sigma  + {\sigma _0}} \right) - 1} \right)\left( {1 + 3\tanh \left( {2\sigma  + {\sigma _0}} \right)} \right),\\
{{\widetilde{\mathcal D}}_ \pm } &= i{\partial _\sigma }{\tau _1} - \left[ {i{\partial _\tau } \mp \frac{1}{2}\left( {1 - \tanh \left( {2\sigma  + {\sigma _0}} \right)} \right)} \right]{\tau _2}\nonumber \\
&\quad+ \frac{1}{{\Omega\, {{\sinh }^2}\sigma }} \mp \frac{1}{{\Omega\, {{\cosh }^2}\left( {\sigma  + {\sigma _0}} \right)}},
\end{align}
with $\sigma_{0}=\infty$ for the 1/2-BPS circle and $\sigma_{0}=0$ for the 1/4-BPS ``special'' latitude\footnote{The parameter $\sigma_0$ here is related to the angle $\theta_0=\arccos\left(\tanh\sigma_0\right)$ parametrizing the entire family of latitude Wilson loops with $\theta_{0}\in [0,\pi/2]$.}, while tilded and untilded operators are related by
\begin{align}\label{TildedandUntilded}
\mathcal{K} = \frac{1}{{{\Omega ^2}}}\widetilde{\mathcal K}, && \mathcal{D} = \frac{1}{{{\Omega ^{3/2}}}}\widetilde{ \mathcal{D}}\, {\Omega ^{1/2}},
\end{align}
with $\Omega$ denoting the circle and latitude conformal factors in \eqref{metric2}.
In equation \eqref{Sdet}, the operator $\mathcal{K}_{1}$ corresponds to bosonic fluctuations around three directions of $AdS$, $\mathcal{K}_{2}$ to three directions along the $S^{3}$ inside $S^{5}$, $\mathcal{K}_{3\pm}$ result from a mixing of the remaining transversal directions to the worldsheet, while $\mathcal{D}_{\pm}$ come from expanding fermions to second order and gauge fixing $\kappa$-symmetry.

It is easy to check that for the latitude ($\sigma_{0}=0$) $\psi_0$ is indeed a zero mode\footnote{The bosonic operators are read by expanding the string action to second order in the form $\int\Omega^{2} \xi^{\hat{a}}\mathcal{K}\xi^{\hat{a}}d\tau d\sigma$ where $\xi^{\hat{a}}$ are fluctuations in the tangent space, as the later is required by the fermionic part. Note that for the ``special'' latitude expanding the string coordinates $X^{\mu}=X^{\mu}|_{\text{cl}}+{E^{\mu}}_{\hat{a}}|_{\rm{cl}}\xi^{\hat{a}}$ and choosing a convenient representation for the vector $\bf{k}$ in \eqref{md1} leads to a vielbein ${E^{\mu}}_{\hat{a}}|_{\rm{cl}}\propto \coth\sigma$ along the $S^{3}$ angles. Along these directions a fluctuation $\xi^{\hat{a}}\propto\tanh\sigma$ amounts to shifting by a constant the values of the angles in the 3-parametric degenerate classical solution.}
\begin{align}
\mathcal{K}_{2}^{\rm{L}}\psi_{0}=\widetilde{\mathcal{K}}_{2}^{\rm{L}}\psi_{0}=0.
\end{align}
\indent Evaluation of the determinants in \eqref{Sdet} is simplified by Fourier expanding along the $\tau$ direction ($i\partial_{\tau}\to\omega$) as the resulting differential operators depend exclusively on $\sigma$. Frequencies are $\omega\in\mathbb{Z}$ for bosons and $\omega\in\mathbb{Z}+1/2$ for fermions in order to account for periodic and anti-periodic boundary conditions.

%%%%%%%%%%%%%%%%%%%%%%%%%%%%%%%%%%%%%%%%%%%%%%%%%%%%%%%%%%%%%%%%%%%%

\subsection{Conformal factors and regulators}\label{ConformalSection}

In the Gel'fand-Yaglom method, just as for phaseshifts, evaluation of determinants requires dropping the conformal factors in the operators $\mathcal{K}$ and $\mathcal{D}$ in \eqref{TildedandUntilded}. The later is achieved by performing a conformal transformation which allows to go from the worldsheet metric \eqref{metric1} to the flat metric $ds^{2}=d\tau^{2}+d\sigma^{2}$. This transformation is singular at $\sigma=\infty$ and changes the topology of the worldsheet from disk to cylinder, but allows to instead consider the determinants of the tilded operators $\widetilde{\mathcal{K}}$ and $\widetilde{\mathcal{D}}$. 

Evaluation of determinants on the cylinder requires the introduction of an IR regulator $R$ at very large $\sigma$. Such procedure potentially breaks diffeomorphism invariance. To circumvent this issue, in \cite{Cagnazzo:2017sny} it was proposed to use different regulators for each Wilson loop such that the area removed by the cutoff is the same. For the present case this results in introducing two IR regulators $R_{\rm{C}}$ and $R_{\rm{L}}$ defined through
$$R\left(\theta_0\right)=R_{\text{inv}}-\frac{1}{2}\ln\frac{1+\cos\theta_{0}}{2},$$
where $R_{\rm{C}}$ and $R_{\rm{L}}$ follow from replacing their corresponding values $\theta_{0}=0$ and $\theta_{0}=\pi/2$, while $R_{\text{inv}}\gg1$ is fixed as it does not depend on $\theta_{0}$.

The 1-loop determinants in the partition function depend exponentially on the IR regulators and a finite remnant remains \cite{Cagnazzo:2017sny}. Performing this conformal transformation from untilded to tilded operators results in
\begin{align}
\frac{Z_{\rm{C}}}{Z_{\rm{L}}}&=e^{\sqrt{\lambda}}\lambda^{-3/4}4\pi\frac{\rm{Sdet'}^{1/2}\widetilde{\mathbb{K}}_{\rm{L}}}{\rm{Sdet}^{1/2}\widetilde{\mathbb{K}}_{\rm{C}}}e^{R_{\rm{C}}-R_{\rm{L}}}\left<\psi_0|\psi_0\right>^{-3/2}\nonumber\\
&=e^{\sqrt{\lambda}}\lambda^{-3/4}\sqrt{\frac{2}{\pi}}\frac{\rm{Sdet'}^{1/2}\widetilde{\mathbb{K}}_{\rm{L}}}{\rm{Sdet}^{1/2}\widetilde{\mathbb{K}}_{\rm{C}}}\frac{e^{R_{\rm{C}}-R_{\rm{L}}}}{\left(\psi_{0}|\psi_{0}\right)^{3/2}},\label{temp1}
\end{align}
with $\left<\psi_0|\psi_0\right>$ denoting the norm of the zero mode on the cylinder
\begin{align}
\left<\psi_0|\psi_0\right>&=\int_{0}^{2\pi}\int_{0}^{R_{\rm{L}}}\tanh^{2}\sigma\ d\sigma d\tau\overset{R_{\rm{L}}\gg1}{\approx}2\pi\left(\psi_0|\psi_0\right),\label{inner}
\end{align}
where we also defined $\left(\psi_{0}|\psi_{0}\right)=\int_{0}^{R_{\rm{inv}}}d\sigma\psi_{0}^{2}$.

Note that the factors at the left of the ratio of determinants in \eqref{temp1} precisely correspond to the field theory prediction \eqref{qftpredictions}.

%%%%%%%%%%%%%%%%%%%%%%%%%%%%%%%%%%%%%%%%%%%%%%%%%%%%%%%%%%%%%%%%%%%%
%%%%%%%%%%%%%%%%%%%%%%%%%%%%%%%%%%%%%%%%%%%%%%%%%%%%%%%%%%%%%%%%%%%%
%%%%%%%%%%%%%%%%%%%%%%%%%%%%%%%%%%%%%%%%%%%%%%%%%%%%%%%%%%%%%%%%%%%%
\section{Evaluation of the functional determinants}\label{Sec4}

\subsection{Determinants with the Gel'fand-Yaglom method}

For all the operators without zero modes, the functional determinants are obtained by using the typical Gel'fand-Yaglom formalism. Note that it is necessary to choose adequate boundary conditions for the evaluation of determinants. Dirichlet-Dirichlet (D-D) boundary conditions are imposed on the operators without zero modes, while the zero mode of $\mathcal{K}_{2}$ will require special consideration. Since the remnant from the conformal transformation from disk to cylinder has already been explicitly included in \eqref{temp1}, the Gel'fand-Yaglom procedure with regulator $R_{\rm{inv}}$ is used for both loops.

In this setup the determinants for the operators in the ratio of 1/2-BPS circular and 1/4-BPS latitude with arbitrary angle $\theta_0$ were computed using Gel'fand-Yaglom in \cite{Forini:2015bgo,Faraggi:2016ekd} and the main results are summarized in \ref{GYAppendex}. The results for the ``special'' latitude follow by taking $\theta_0=\frac{\pi}{2}$ in equations \eqref{A.1}-\eqref{A.4}, except for the operator $\widetilde{\mathcal{K}}_2$ with $\omega=0$ where the zero mode exists. We obtain 
\begin{align}
\ln \frac{{{\rm{det}}{\kern 1pt} {{\widetilde{\mathcal K}}_{1}^{\text{L}}}}}{{{\rm{det}}{\kern 1pt} {{\widetilde{\mathcal K}}_{1}^{\text{C}}}}} &= 0,\label{4.1}\\
\ln \frac{{\det' {{\widetilde{\mathcal K}}_{2}^{\text{L}}}}}{{\det {\kern 1pt} {{\widetilde{\mathcal K}}_{2}^{\text{C}}}}} &=\sum\limits_{\substack{\omega  \in \mathbb{Z}\\ \omega\neq0}} {\ln \frac{{\left| \omega  \right|}}{{\left| \omega  \right| + 1}}}+\ln\mathcal{P}_0\nonumber\\
&=2\mathcal{F}(\Lambda)+\ln\mathcal{P}_0,\label{4.2}\\
\ln \frac{{\det {{\widetilde{\mathcal K}}_{3\pm}^{\text{L}}}}}{{\det {\kern 1pt} {{\widetilde{\mathcal K}}_{3\pm}^{\text{C}}}}} &= \frac{1}{2}\sum\limits_{\omega  \in \mathbb{Z}} {\ln \frac{{\left| \omega  \right| + 1 }}{{\left| \omega  \right| + 2}}}\nonumber\\
&=\mathcal{F}(\Lambda)-\frac{1}{2}\ln\frac{1}{2},\label{4.3}\\
\ln \frac{{\det {\mathcal{\widetilde D}_{\pm}^{\text{L}} }}}{{\det {\mathcal{\widetilde D}_{\pm}^{\text{C}} }}} &= \frac{1}{2}\sum\limits_{\omega  \in \mathbb{Z}+1/2} {\ln \frac{{\left| {\omega} \right| + \frac{1}{2}}}{{\left| {\omega } \right| + \frac{3}{2}}}}\nonumber\\
&=\mathcal{F}(\Lambda),\label{4.4}
\end{align}
where $\mathcal{P}_0$ denotes the ratio of the functional determinants of $\widetilde{\mathcal{K}}_2$ with $\omega=0$, and $\mathcal{F}(\Lambda)$ is defined by 
\begin{equation}
\mathcal{F}(\Lambda)=\sum_{\omega=1}^{\Lambda}\ln\frac{|\omega|}{|\omega|+1}
\end{equation}
with a symmetric regulator denoted by $\Lambda$. Taking the limit $\Lambda\to\infty$ will recover the original summations. Collecting all the pieces in \eqref{4.1}-\eqref{4.4},  one finds
\begin{equation}
\frac{\rm{Sdet'}^{1/2}\widetilde{\mathbb{K}}_{\text{L}}}{\rm{Sdet}^{1/2}\widetilde{\mathbb{K}}_{\text{C}}}=\sqrt{2}\ \mathcal{P}_0^{{3}/{2}}.
\end{equation}
The numerical factor in front of $\mathcal{P}_0$ above cancels the remnant of the conformal map from section \ref{ConformalSection}, resulting in
\begin{equation}
\frac{Z_{\rm{C}}}{Z_{\rm{L}}}=e^{\sqrt{\lambda}}\lambda^{-3/4}\sqrt{\frac{2}{\pi}}\frac{\mathcal{P}_0^{3/2}}{\left(\psi_0|\psi_0\right)^{3/2}}.\label{30}
\end{equation}
To obtain the final result we now focus on the operators with zero modes.

%%%%%%%%%%%%%%%%%%%%%%%%%%%%%%%%%%%%%%%%%%%%%%%%%%%%%%%%%%%%%%%%%%%%

\subsection{Excluding the zero mode}\label{ZeroSection}
A naive application of the Gel'fand-Yaglom formalism on $\mathcal{P}_0$ will result in a vanishing partition function. Thus, it is necessary to omit the zero mode. To achieve this goal, two modifications are expected. First, the D-D boundary condition should be replaced by a Dirichlet-Neumann (D-N) boundary condition since the zero modes will be ruled out by the first one. Second, the Gel'fand-Yaglom formalism should be modified so that it is suitable for the operators with zero modes. In the following subsections, first we introduce a modified Gel'fand-Yaglom formula, then a short derivation is provided, and finally we apply it to $\mathcal{P}_0$.
\subsubsection{The modified Gel'fand-Yaglom formula}
Consider two second order differential operators defined by
\begin{align}
\widetilde{\mathcal{K}}^{\rm{L}}=-\partial_{\sigma}^2-V(\sigma),\qquad\widetilde{\mathcal{K}}^{\rm{C}}=-\partial_{\sigma}^2,\label{32}
\end{align}
where $V(\sigma)$ is an arbitrary potential with the limit $V(\sigma\to\infty)=0$. We assume that there is only one zero mode for $\widetilde{\mathcal{K}}^{\rm{L}}$. Under the D-N boundary condition, the modified Gel'fand-Yaglom formula for such operators is\footnote{A similar expression for a modified Gel'fand-Yaglom formula under the D-D boundary condition has been given in \cite{falco2017functional,KIRSTEN2003502,Kirsten:2005di}.}
\begin{equation}
\frac{\det'\widetilde{\mathcal{K}}^{\rm{L}}}{\det\widetilde{\mathcal{K}}^{\rm{C}}}=\frac{\left(\phi_{0}^{\rm{L}}|\phi_{0}^{\rm{L}}\right)}{\phi_{0}^{\rm{L}}(R_{\rm{inv}})\partial_{\sigma}\phi^{\rm{C}}_0(R_{\rm{inv}})},\label{mgy}
\end{equation}
where $\phi_0^{\rm{L}}$ and $\phi_0^{\rm{C}}$ are the eigenfunctions of $\widetilde{\mathcal{K}}^{\rm{L},\rm{C}}$ with eigenvalue zero satisfying the boundary conditions\footnote{For an arbitrary second order differential operator $\widetilde{\mathcal{K}}^{\rm{L}}$ without zero modes, the usual Gel'fand-Yaglom formula with the D-N boundary condition is $\det\widetilde{\mathcal{K}}^{\rm{L}}/\det\widetilde{\mathcal{K}}^{\rm{C}}=\partial_{\sigma}\phi^{\rm{L}}(R_{\rm{inv}})/\partial_{\sigma}\phi^{\rm{C}}(R_{\rm{inv}})$, where $\phi^{\rm{C,L}}(\sigma)$ is the solution to $\widetilde{\mathcal{K}}^{\rm{C,L}}\phi^{\rm{C,L}}(\sigma)=0$ satisfying \eqref{boundary}.\label{fn7}}
\begin{align}
\phi_0^{\rm{C,L}}(0)=0, && \partial_{\sigma}\phi_{0}^{\rm{C,L}}(0)=1,
\label{boundary}
\end{align}
while the inner product in the numerator of \eqref{mgy} follows the same definition as the one for $\psi_0$ in (\ref{inner}). In the next section we provide a derivation for \eqref{mgy}.
\subsubsection{Derivation of the modified Gel'fand-Yaglom formula}
Gel'fand-Yaglom formulas for determinants of differential operators with zero modes can be obtained using different approaches, for instance, using the Wronski construction of Green functions \cite{Kleinert:1998rz,kleinert1999functional}, modifying slightly the operator or using the results of \cite{McKane:1995vp} where the boundary conditions are perturbed. A simpler route is perhaps to use as starting point the contour integration formalism proposed in \cite{KIRSTEN2003502,Kirsten:2005di} for the case of D-D boundary conditions. Modifying this construction for the case of D-N boundary conditions leads to \eqref{mgy}. The main idea of this procedure is to construct a function which vanishes at all non-zero eingenvalues, but not at the zero eigenvalue. Using such a function it is easy to construct a modified zeta function for the operator and obtain its determinant. 

%There are many routes leading to \eqref{mgy}. However it will be inspirational to follow the procedure starting from the zeta function regularization\footnote{The procedure is motivated by the calculation in \cite{KIRSTEN2003502}, which provided a detailed derivation for the formalism under D-D boundary condition but not D-N boundary condition.}.

The zeta function of an arbitrary differential operator $\widetilde{\mathcal{K}}^{\rm{L}}$ with specified boundary conditions (D-N in our case) is defined by $\zeta_{\widetilde{\mathcal{K}}^{\rm{L}}}(s)=\sum_{n}\lambda^{-s}_{n}$ where $\text{Re}(s)>1$ and $\lambda_n$ are the eigenvalues of the operator which are assumed to be non-negative. By analytically continuing the zeta function to $s=0$, the functional determinant of the operator is related to it by $\ln\det\widetilde{\mathcal{K}}^{\rm{L}}=-\zeta_{\widetilde{\mathcal{K}}^{\rm{L}}}'(0)$. 

We denote by $\phi_{\beta}^{\rm{L}}(\sigma)$ the solutions to $\widetilde{\mathcal{K}}^{\rm{L}}\phi_{\beta}^{\rm{L}}(\sigma)=\beta^2\phi_{\beta}^{\rm{L}}(\sigma)$ with boundary conditions $\phi_{\beta}^{\rm{L}}(0)=0$ and $\partial_{\sigma}\phi_{\beta}^{\rm{L}}(0)=1$. The first one imposes the Dirichlet boundary condition, while the second fixes the normalization. The eigenvalues $\lambda_n$ are determined by the Neumann boundary condition $\partial_{\sigma}\phi_{\lambda_{n}}^{\rm{L}}(R_{\rm{inv}})=0.$ Thus, the zeta function of $\widetilde{\mathcal{K}}^{\rm{L}}$ can be written in the form of a contour integral
\begin{equation}
\zeta_{\widetilde{\mathcal{K}}^{\rm{L}}}(s)=\frac{1}{2\pi i}\oint d\beta\ \beta^{-2s}\partial_{\beta}\ln\partial_{\sigma}\phi_{\beta}^{\rm{L}}(R_{\rm{inv}}),\label{zeta}
\end{equation}
with the contour surrounding the whole non-negative real axis. The above expression is actually divergent since the zero eigenvalue is also included, so it is necessary to study the asymptotic behaviour of the zero modes for the purpose of omitting the zero eigenvalue. Now consider the equation
\begin{equation}
\int_0^{R_{\rm{inv}}}d\sigma\phi_{0}^{\rm{L}}\widetilde{\mathcal{K}}^{\rm{L}}\phi_{\beta}^{\rm{L}}=\beta^2\int_0^{R_{\rm{inv}}}d\sigma\phi_{0}^{\rm{L}}\phi_{\beta}^{\rm{L}}.\label{37}
\end{equation}
Integrating the left hand side by parts twice moves the operator $\widetilde{\mathcal{K}}^{\rm{L}}$ to the left of $\phi_0^{\rm{L}}$ which results in zero, and produces two boundary terms. The integral on the right hand side is the inner product $\left(\phi_{0}^{\rm{L}}|\phi_{\beta}^{\rm{L}}\right)$. Applying the boundary conditions, one finds
\begin{equation}
\partial_{\sigma}\phi_{\beta}^{\rm{L}}(R_{\rm{inv}})=-\frac{\beta^2\left(\phi_{0}^{\rm{L}}|\phi_{\beta}^{\rm{L}}\right)}{\phi_{0}^{\rm{L}}(R_{\rm{inv}})}\equiv-\beta^2u_{\beta}(R_{\rm{inv}}).\label{v}
\end{equation}
Note that $u_{\lambda_n}(R_{\rm{inv}})=0$ as long as $\lambda_n\neq0$, and $u_{\lambda_n}(R_{\rm{inv}})\neq0$ when $\lambda_n=0$ since the inner product $\left(\phi_{0}^{\rm{L}}|\phi_{0}^{\rm{L}}\right)$ is positive definite.
%where the left hand side has roots at the eigenvalues satisfying D-N boundary conditions. Note that $u_{\beta}(R_{\rm{inv}})=0$ as long as $\beta\neq0$, and $u_{\beta}(R_{\rm{inv}})\neq0$ when $\beta=0$ since the inner product $\left(\phi_{0}^{\rm{L}}|\phi_{0}^{\rm{L}}\right)$ is positive definite. 
Thus, the roots of $u_{\beta}(R_{\rm{inv}})=0$ build out the whole non-vanishing operator spectrum. Consequently, it is natural to replace $\partial_{\sigma}\phi_{\beta}^{\rm{L}}(R_{\rm{inv}})$ in \eqref{zeta} by $u_{\beta}(R_{\rm{inv}})$ to remove the zero in the spectrum. To preserve the asymptotic behaviour of $\partial_{\sigma}\phi_{\beta}^{\rm{L}}(R_{\rm{inv}})$ as $\beta\to\infty$, we define a function $v_{\beta}(R_{\rm{inv}})\equiv(1-\beta^2)u_{\beta}(R_{\rm{inv}})$. The zeta function, omitting the zero eigenvalue, goes as 
\begin{align}
\zeta_{\widetilde{\mathcal{K}}^{\rm{L}}}(s)%=&\frac{1}{2\pi i}\oint d\beta\  \beta^{-2s}\partial_{\beta}\ln u_{\beta}(R_{\rm{L}})\nonumber \\
&=\frac{1}{2\pi i}\oint d\beta\  \beta^{-2s}\left(\partial_{\beta}\ln v_{\beta}(R_{\rm{inv}})\nonumber-\partial_{\beta}\ln (1-\beta^2)\right).
\end{align}
Note that the contour integral of the second term just equals to one. To evaluate the integral of the first term, one can deform the contour to the imaginary axis by giving $\beta$ in the upper (lower) part of the contour a phase $e^{i\pi/2}$ $(e^{-i\pi/2})$. Since there are no poles on the imaginary axis, the integration equals to 
\begin{align}
\zeta_{\widetilde{\mathcal{K}}^{\rm{L}}}(s)=\frac{\sin{\pi s}}{\pi}\int_0^{\infty}d\beta\ \beta^{-2s}\left(\partial_{\beta}\ln v_{i\beta}(R_{\rm{inv}})\right)-1.
\end{align}
The resulting functional determinant omitting the zero mode is
\begin{align}
\ln{\rm{det'}}\widetilde{\mathcal{K}}^{\rm{L}}=-\zeta'_{\widetilde{\mathcal{K}}^{\rm{L}}}(0)=-\ln v_{i\beta}(R_{\rm{inv}})|_{\beta=0}^{\infty}=\ln\frac{u_{0}(R_{\rm{inv}})}{\partial_{\sigma}\phi_{i\infty}^{\rm{L}}(R_{\rm{inv}})},
\end{align} 
where we used \eqref{v} on the right hand side.

On the other hand, there is no zero mode for $\widetilde{\mathcal{K}}^{\rm{C}}$, so its zeta function is defined in the typical way by \eqref{zeta} with $\partial_{\sigma}\phi^{\rm{C}}_{\beta}(R_{\rm{inv}})$, which results in\footnote{The Gel'fand-Yaglom formula for the case of D-N boundary conditions without zero modes shown in footnote \ref{fn7} follows directly from \eqref{nozero}.} 
\begin{align}
\ln\det\widetilde{\mathcal{K}}^{\rm{C}}=\ln\frac{\partial_{\sigma}\phi_{0}^{\rm{C}}(R_{\rm{inv}})}{\partial_{\sigma}\phi_{i\infty}^{\rm{C}}(R_{\rm{inv}})}\label{nozero}.
\end{align}
Notice that in the limit of $|\beta|\to\infty$, $\phi_{\beta}^{\rm{L}}=\phi^{\rm{C}}_{\beta}$, one finally obtains
\begin{equation}
\begin{split}
\ln\frac{{\rm{det'} {\widetilde{\mathcal{K}}^{\rm{L}}}}}{{\det {\kern 1pt} {{\widetilde{\mathcal{K}}}^{\rm{C}}}}}=\ln\frac{u_{0}(R_{\rm{inv}})}{\partial_{\sigma}\phi^{\rm{ C}}_0(R_{\rm{inv}})}=\ln\frac{\left(\phi_{0}^{\rm{L}}|\phi_{0}^{\rm{L}}\right)}{\phi_{0}^{\rm{L}}(R_{\rm{inv}})\partial_{\sigma}\phi^{\rm{C}}_0(R_{\rm{inv}})}.
\end{split}
\end{equation}
\subsubsection{Using the modified Gel'fand-Yaglom formula}
To calculate $\mathcal{P}_0$, we take $\widetilde{\mathcal{K}}^{\rm{L}}\to\widetilde{\mathcal{K}}_{2}^{\rm{L}}|_{\omega=0},\  \widetilde{\mathcal{K}}^{\rm{C}}\to\widetilde{\mathcal{K}}_{2}^{\rm{C}}|_{\omega=0}.$
By solving the zero eigenvalue equations for the differential operators subjected to the boundary conditions \eqref{boundary}, one finds
\begin{align}
\phi^{\rm{C}}_0(\sigma)=\sigma,\qquad\qquad \phi_{0}^{\rm{L}}(\sigma)=\tanh\sigma=\psi_{0}.
\end{align}
Following the recipe in \eqref{mgy} with $R_{\rm{inv}}\to\infty$, 
\begin{align}
\mathcal{P}_0=\frac{\det'\widetilde{\mathcal{K}}_{2}^{\rm{L}}|_{\omega=0}}{\det\widetilde{\mathcal{K}}_{2}^{\rm{C}}|_{\omega=0}}=\left(\psi_{0}|\psi_{0}\right).\label{36}
\end{align}
Combining \eqref{30} and \eqref{36} retrieves the exact gauge theory prediction\footnote{Notice that there is an explicit cancellation of the $\left(\psi_{0}|\psi_{0}\right)$ factors from $\rm{Sdet'}\mathbb{K}_{\rm{L}}$ and the Jacobian from the transformation to collective coordinates. The later is a phenomenon observed in instanton methods and could be used to greatly simplify calculations \cite{McKane:1995vp,falco2017functional}.}
\begin{equation}
\frac{W_{\rm{C}}}{W_{\rm{L}}}=\frac{Z_{\rm{C}}}{Z_{\rm{L}}}=e^{\sqrt{\lambda}}\lambda^{-3/4}\sqrt{\frac{2}{\pi}}.
\end{equation}

%%%%%%%%%%%%%%%%%%%%%%%%%%%%%%%%%%%%%%%%%%%%%%%%%%%%%%%%%%%%%%%%%%%%
%%%%%%%%%%%%%%%%%%%%%%%%%%%%%%%%%%%%%%%%%%%%%%%%%%%%%%%%%%%%%%%%%%%%
%%%%%%%%%%%%%%%%%%%%%%%%%%%%%%%%%%%%%%%%%%%%%%%%%%%%%%%%%%%%%%%%%%%%

\section{Conclusions}\label{Sec5}
In this letter we have revisited the computation of the 1/2-BPS circular Wilson at 1-loop in $AdS_5\times S^5$ in the setup of \cite{Medina-Rincon:2018wjs} by using Gel'fand-Yaglom results and the zero mode exclusion method of \cite{falco2017functional,KIRSTEN2003502}. As in the phaseshift computation, exact matching with the field theory prediction is achieved and the final result follows from the zero modes of the latitude moduli as other contributions cancel. The later is one of several similarities between the phaseshift and Gel'fand-Yaglom techniques, which are among the most popular methods for precision holography for Wilson loops having achieved similar success for $AdS_4\times CP^{3}$ \cite{Medina-Rincon:2019bcc,David:2019lhr}. It would be interesting to elaborate on the possible connections between these methods in further works\footnote{Connections between the phaseshift and Heat kernel methods have been recently discussed in \cite{1805525}.}.

Another string Wilson loop calculation where the techniques used here for zero modes could find application is the ratio of fermionic 1/2-BPS and bosonic 1/6-BPS circular Wilson loops in ABJM \cite{Drukker:2019bev}. According to localization the ratio of these Wilson loops is proportional to ${\lambda}^{-1/2}$ at large $\lambda$ \cite{Drukker:2010nc}. In this setup the string solution for the bosonic circular Wilson loop in $AdS_{4}\times CP^{3}$ is expected to have two zero modes related to the smearing over a $CP^{1}$ inside $CP^{3}$ \cite{Drukker:2008zx}.

Zero modes play an important role when studying the contributions of Fadeev-Popov ghosts and bosonic longitudinal modes, an issue that has held back the calculation of individual Wilson loops in string theory. It would be interesting to see if extensions of techniques like the one used here, along with a better understanding of the required boundary conditions, could shed some light into this problem.

\section*{Acknowledgements}
We would like to thank K. Zarembo for comments on the manuscript. This work was partially supported by the Swiss National Science Foundation through the NCCR SwissMAP. The work of L.B. was partially supported by the Scholarship for international MSc students from ETH Zurich.

%%%%%%%%%%%%%%%%%%%%%%%%%%%%%%%%%%%%%%%%%%%%%%%%%%%%%%%%%%%%%%%%%%%%
%%%%%%%%%%%%%%%%%%%%%%%%%%%%%%%%%%%%%%%%%%%%%%%%%%%%%%%%%%%%%%%%%%%%
%%%%%%%%%%%%%%%%%%%%%%%%%%%%%%%%%%%%%%%%%%%%%%%%%%%%%%%%%%%%%%%%%%%%

\appendix
\section{Gel'fand-Yaglom results for latitude Wilson loop determinants}\label{GYAppendex}
The evaluation of functional determinants for the ratio of a 1/2-BPS circle and 1/4-BPS latitude Wilson loop with arbitrary angle $\theta_0\in[0,\pi/2)$ was carried out in  \cite{Forini:2015bgo,Faraggi:2016ekd} using Gel'fand-Yaglom obtaining
\begin{align}
\ln \frac{{{\rm{det}}{\kern 1pt} {{\widetilde{\mathcal K}}_1}\left( {{\theta _0}} \right)}}{{{\rm{det}}{\kern 1pt} {{\widetilde{\mathcal K}}_1}\left( 0 \right)}} &= 0,\label{A.1}\\
\ln \frac{{\det {{\widetilde{\mathcal K}}_2}\left( {{\theta _0}} \right)}}{{\det {\kern 1pt} {{\widetilde{\mathcal K}}_2}\left( 0 \right)}} &=\sum\limits_{\omega  \in \mathbb{Z}} {\ln \frac{{\left| \omega  \right| + \cos {\theta _0}}}{{\left| \omega  \right| + 1}}},\label{A.2}\\
\ln \frac{{\det {{\widetilde{\mathcal K}}_{3\pm}}\left( {{\theta _0}} \right)}}{{\det {\kern 1pt} {{\widetilde{\mathcal K}}_{3\pm}}\left( 0 \right)}} &= \frac{1}{2}\sum\limits_{\omega  \in \mathbb{Z}} {\ln \frac{{\left| \omega  \right| + 1+\cos\theta_0 }}{{\left| \omega  \right| + 2}}} ,\label{A.3}\\
\ln \frac{{\det {\mathcal{\widetilde D}_\pm }\left( {{\theta _0}} \right)}}{{\det {\mathcal{\widetilde D}_\pm }\left( 0 \right)}} &= \frac{1}{2}\sum\limits_{\omega  \in \mathbb{Z}+1/2} {\ln \frac{{\left| {\omega} \right| + \frac{1}{2} + \cos {\theta _0}}}{{\left| {\omega } \right| + \frac{3}{2}}}}\label{A.4} .
\end{align}
Using these results in \eqref{Sdet} leads to the main result of \cite{Forini:2015bgo,Faraggi:2016ekd}
\begin{align}
\ln\frac{\rm{Sdet}^{1/2}\mathbb{K}(\theta_0) }{\rm{Sdet}^{1/2}\mathbb{K}(0) }= \frac{3}{2}\ln \cos {\theta _0} -\frac{1}{2} \ln \frac{{1 + \cos {\theta _0}}}{2}.
\end{align}
The second term on the right is cancelled by careful consideration of the mapping from the disk to the cylinder as explained in section \ref{ConformalSection} and \cite{Cagnazzo:2017sny}, leading to the field theory prediction. Note that the final answer comes exclusively from the $\omega=0$ eigenvalues of $\widetilde{\mathcal{K}}_{2}$ as all the other contributions cancel.

These results are carried on for the case considered here where $\theta_0=\pi/2$, except for the zero mode of $\widetilde{\mathcal{K}}_2$ for which $\omega=0$. The later requires special treatment as explained in section \ref{ZeroSection}.

%%%%%%%%%%%%%%%%%%%%%%%%%%%%%%%%%%%%%%%%%%%%%%%%%%%%%%%%%%%%%%%%%%%%
%%%%%%%%%%%%%%%%%%%%%%%%%%%%%%%%%%%%%%%%%%%%%%%%%%%%%%%%%%%%%%%%%%%%
%%%%%%%%%%%%%%%%%%%%%%%%%%%%%%%%%%%%%%%%%%%%%%%%%%%%%%%%%%%%%%%%%%%%

\bibliographystyle{elsarticle-num}

\bibliography{refs}

\begin{thebibliography}{10}
\expandafter\ifx\csname url\endcsname\relax
  \def\url#1{\texttt{#1}}\fi
\expandafter\ifx\csname urlprefix\endcsname\relax\def\urlprefix{URL }\fi
\expandafter\ifx\csname href\endcsname\relax
  \def\href#1#2{#2} \def\path#1{#1}\fi

\bibitem{Maldacena:1997re}
J.~M. Maldacena, {The Large N limit of superconformal field theories and
  supergravity}, Int. J. Theor. Phys. 38 (1999) 1113--1133, [Adv. Theor. Math.
  Phys.2,231(1998)].
\newblock \href {http://arxiv.org/abs/hep-th/9711200}
  {\path{arXiv:hep-th/9711200}}, \href
  {https://doi.org/10.1023/A:1026654312961, 10.4310/ATMP.1998.v2.n2.a1}
  {\path{doi:10.1023/A:1026654312961, 10.4310/ATMP.1998.v2.n2.a1}}.

\bibitem{Maldacena:1998im}
J.~M. Maldacena, {Wilson loops in large N field theories}, Phys. Rev. Lett. 80
  (1998) 4859--4862.
\newblock \href {http://arxiv.org/abs/hep-th/9803002}
  {\path{arXiv:hep-th/9803002}}, \href
  {https://doi.org/10.1103/PhysRevLett.80.4859}
  {\path{doi:10.1103/PhysRevLett.80.4859}}.

\bibitem{Rey:1998ik}
S.-J. Rey, J.-T. Yee, {Macroscopic strings as heavy quarks in large N gauge
  theory and anti-de Sitter supergravity}, Eur. Phys. J. C 22 (2001) 379--394.
\newblock \href {http://arxiv.org/abs/hep-th/9803001}
  {\path{arXiv:hep-th/9803001}}, \href {https://doi.org/10.1007/s100520100799}
  {\path{doi:10.1007/s100520100799}}.

\bibitem{Erickson:2000af}
J.~K. Erickson, G.~W. Semenoff, K.~Zarembo, {Wilson loops in N=4 supersymmetric
  Yang-Mills theory}, Nucl. Phys. B582 (2000) 155--175.
\newblock \href {http://arxiv.org/abs/hep-th/0003055}
  {\path{arXiv:hep-th/0003055}}, \href
  {https://doi.org/10.1016/S0550-3213(00)00300-X}
  {\path{doi:10.1016/S0550-3213(00)00300-X}}.

\bibitem{Pestun:2007rz}
V.~Pestun, {Localization of gauge theory on a four-sphere and supersymmetric
  Wilson loops}, Commun. Math. Phys. 313 (2012) 71--129.
\newblock \href {http://arxiv.org/abs/0712.2824} {\path{arXiv:0712.2824}},
  \href {https://doi.org/10.1007/s00220-012-1485-0}
  {\path{doi:10.1007/s00220-012-1485-0}}.

\bibitem{Polyakov:1997tj}
A.~M. Polyakov, {String theory and quark confinement}, Nucl. Phys. Proc. Suppl.
  68 (1998) 1--8.
\newblock \href {http://arxiv.org/abs/hep-th/9711002}
  {\path{arXiv:hep-th/9711002}}, \href
  {https://doi.org/10.1016/S0920-5632(98)00135-2}
  {\path{doi:10.1016/S0920-5632(98)00135-2}}.

\bibitem{Drukker:1999zq}
N.~Drukker, D.~J. Gross, H.~Ooguri, {Wilson loops and minimal surfaces}, Phys.
  Rev. D 60 (1999) 125006.
\newblock \href {http://arxiv.org/abs/hep-th/9904191}
  {\path{arXiv:hep-th/9904191}}, \href
  {https://doi.org/10.1103/PhysRevD.60.125006}
  {\path{doi:10.1103/PhysRevD.60.125006}}.

\bibitem{Drukker:2000ep}
N.~Drukker, D.~J. Gross, A.~A. Tseytlin, {Green-Schwarz string in AdS(5) x
  S**5: Semiclassical partition function}, JHEP 04 (2000) 021.
\newblock \href {http://arxiv.org/abs/hep-th/0001204}
  {\path{arXiv:hep-th/0001204}}, \href
  {https://doi.org/10.1088/1126-6708/2000/04/021}
  {\path{doi:10.1088/1126-6708/2000/04/021}}.

\bibitem{Kruczenski:2008zk}
M.~Kruczenski, A.~Tirziu, {Matching the circular Wilson loop with dual open
  string solution at 1-loop in strong coupling}, JHEP 05 (2008) 064.
\newblock \href {http://arxiv.org/abs/0803.0315} {\path{arXiv:0803.0315}},
  \href {https://doi.org/10.1088/1126-6708/2008/05/064}
  {\path{doi:10.1088/1126-6708/2008/05/064}}.

\bibitem{Kristjansen:2012nz}
C.~Kristjansen, Y.~Makeenko, {More about One-Loop Effective Action of Open
  Superstring in $AdS_5\times S^5$}, JHEP 09 (2012) 053.
\newblock \href {http://arxiv.org/abs/1206.5660} {\path{arXiv:1206.5660}},
  \href {https://doi.org/10.1007/JHEP09(2012)053}
  {\path{doi:10.1007/JHEP09(2012)053}}.

\bibitem{Buchbinder:2014nia}
E.~I. Buchbinder, A.~A. Tseytlin, {1/N correction in the D3-brane description
  of a circular Wilson loop at strong coupling}, Phys. Rev. D89~(12) (2014)
  126008.
\newblock \href {http://arxiv.org/abs/1404.4952} {\path{arXiv:1404.4952}},
  \href {https://doi.org/10.1103/PhysRevD.89.126008}
  {\path{doi:10.1103/PhysRevD.89.126008}}.

\bibitem{Giombi:2020mhz}
S.~Giombi, A.~A. Tseytlin, {Strong coupling expansion of circular Wilson loops
  and string theories in AdS$_5 \times S^5$ and AdS$_4 \times CP^3$} (7 2020).
\newblock \href {http://arxiv.org/abs/2007.08512} {\path{arXiv:2007.08512}}.

\bibitem{Forini:2015bgo}
V.~Forini, V.~Giangreco M.~Puletti, L.~Griguolo, D.~Seminara, E.~Vescovi,
  {Precision calculation of 1/4-BPS Wilson loops in AdS$_5\times S^5$}, JHEP 02
  (2016) 105.
\newblock \href {http://arxiv.org/abs/1512.00841} {\path{arXiv:1512.00841}},
  \href {https://doi.org/10.1007/JHEP02(2016)105}
  {\path{doi:10.1007/JHEP02(2016)105}}.

\bibitem{Faraggi:2016ekd}
A.~Faraggi, L.~A. Pando~Zayas, G.~A. Silva, D.~Trancanelli, {Toward precision
  holography with supersymmetric Wilson loops}, JHEP 04 (2016) 053.
\newblock \href {http://arxiv.org/abs/1601.04708} {\path{arXiv:1601.04708}},
  \href {https://doi.org/10.1007/JHEP04(2016)053}
  {\path{doi:10.1007/JHEP04(2016)053}}.

\bibitem{Forini:2017whz}
V.~Forini, A.~A. Tseytlin, E.~Vescovi, {Perturbative computation of string
  one-loop corrections to Wilson loop minimal surfaces in AdS$_5 \times$
  S$^5$}, JHEP 03 (2017) 003.
\newblock \href {http://arxiv.org/abs/1702.02164} {\path{arXiv:1702.02164}},
  \href {https://doi.org/10.1007/JHEP03(2017)003}
  {\path{doi:10.1007/JHEP03(2017)003}}.

\bibitem{Cagnazzo:2017sny}
A.~Cagnazzo, D.~Medina-Rincon, K.~Zarembo, {String corrections to circular
  Wilson loop and anomalies}, JHEP 02 (2018) 120.
\newblock \href {http://arxiv.org/abs/1712.07730} {\path{arXiv:1712.07730}},
  \href {https://doi.org/10.1007/JHEP02(2018)120}
  {\path{doi:10.1007/JHEP02(2018)120}}.

\bibitem{Medina-Rincon:2018wjs}
D.~Medina-Rincon, A.~A. Tseytlin, K.~Zarembo, {Precision matching of circular
  Wilson loops and strings in $AdS_{5}\times S^{5}$}, JHEP 05 (2018) 199.
\newblock \href {http://arxiv.org/abs/1804.08925} {\path{arXiv:1804.08925}},
  \href {https://doi.org/10.1007/JHEP05(2018)199}
  {\path{doi:10.1007/JHEP05(2018)199}}.

\bibitem{Zarembo:2002an}
K.~Zarembo, {Supersymmetric Wilson loops}, Nucl. Phys. B643 (2002) 157--171.
\newblock \href {http://arxiv.org/abs/hep-th/0205160}
  {\path{arXiv:hep-th/0205160}}, \href
  {https://doi.org/10.1016/S0550-3213(02)00693-4}
  {\path{doi:10.1016/S0550-3213(02)00693-4}}.

\bibitem{doi:10.1063/1.1703636}
I.~M. Gel'fand, A.~M. Yaglom, Integration in functional spaces and its
  applications in quantum physics, Journal of Mathematical Physics 1~(1) (1960)
  48--69.
\newblock \href {https://doi.org/10.1063/1.1703636}
  {\path{doi:10.1063/1.1703636}}.

\bibitem{McKane:1995vp}
A.~J. McKane, M.~B. Tarlie, {Regularization of functional determinants using
  boundary perturbations}, J. Phys. A 28 (1995) 6931--6942.
\newblock \href {http://arxiv.org/abs/cond-mat/9509126}
  {\path{arXiv:cond-mat/9509126}}, \href
  {https://doi.org/10.1088/0305-4470/28/23/032}
  {\path{doi:10.1088/0305-4470/28/23/032}}.

\bibitem{Kleinert:1998rz}
H.~Kleinert, A.~Chervyakov, {Simple explicit formulas for Gaussian path
  integrals with time dependent frequencies}, Phys. Lett. A 245 (1998)
  345--357.
\newblock \href {http://arxiv.org/abs/quant-ph/9803016}
  {\path{arXiv:quant-ph/9803016}}, \href
  {https://doi.org/10.1016/S0375-9601(98)00380-6}
  {\path{doi:10.1016/S0375-9601(98)00380-6}}.

\bibitem{kleinert1999functional}
H.~Kleinert, A.~Chervyakov, Functional determinants from wronski green
  functions, Journal of Mathematical Physics 40~(11) (1999) 6044--6051.

\bibitem{falco2017functional}
G.~Falco, A.~A. Fedorenko, I.~A. Gruzberg, On functional determinants of matrix
  differential operators with multiple zero modes, Journal of Physics A:
  Mathematical and Theoretical 50~(48) (2017) 485201.

\bibitem{KIRSTEN2003502}
K.~Kirsten, A.~J. McKane, {Functional determinants by contour integration
  methods}, Annals Phys. 308 (2003) 502--527.
\newblock \href {http://arxiv.org/abs/math-ph/0305010}
  {\path{arXiv:math-ph/0305010}}, \href
  {https://doi.org/10.1016/S0003-4916(03)00149-0}
  {\path{doi:10.1016/S0003-4916(03)00149-0}}.

\bibitem{Kirsten:2005di}
K.~Kirsten, A.~J. McKane, {Functional determinants in the presence of zero
  modes}, in: {6th Workshop on Quantum Field Theory under the Influence of
  External Conditions (QFEXT03)}, 2005, pp. 146--151.
\newblock \href {http://arxiv.org/abs/hep-th/0507005}
  {\path{arXiv:hep-th/0507005}}.

\bibitem{Drukker:2000rr}
N.~Drukker, D.~J. Gross, {An Exact prediction of N=4 SUSYM theory for string
  theory}, J. Math. Phys. 42 (2001) 2896--2914.
\newblock \href {http://arxiv.org/abs/hep-th/0010274}
  {\path{arXiv:hep-th/0010274}}, \href {https://doi.org/10.1063/1.1372177}
  {\path{doi:10.1063/1.1372177}}.

\bibitem{Medina-Rincon:2019bcc}
D.~Medina-Rincon, {Matching quantum string corrections and circular Wilson
  loops in $AdS_4 \times CP^3$}, JHEP 08 (2019) 158.
\newblock \href {http://arxiv.org/abs/1907.02984} {\path{arXiv:1907.02984}},
  \href {https://doi.org/10.1007/JHEP08(2019)158}
  {\path{doi:10.1007/JHEP08(2019)158}}.

\bibitem{David:2019lhr}
M.~David, R.~De~Le\'on~Ard\'on, A.~Faraggi, L.~A. Pando~Zayas, G.~A. Silva,
  {One-loop holography with strings in $AdS_4\times\mathbb {CP}^3$}, JHEP 10
  (2019) 070.
\newblock \href {http://arxiv.org/abs/1907.08590} {\path{arXiv:1907.08590}},
  \href {https://doi.org/10.1007/JHEP10(2019)070}
  {\path{doi:10.1007/JHEP10(2019)070}}.

\bibitem{1805525}
R.~d.~L. Ard\'on, {On the heat kernel and semiclassical $p$-branes in
  hyperbolic space} (7 2020).
\newblock \href {http://arxiv.org/abs/2007.03591} {\path{arXiv:2007.03591}}.

\bibitem{Drukker:2019bev}
N.~Drukker, et~al., {Roadmap on Wilson loops in 3d Chern--Simons-matter
  theories}, J. Phys. A 53~(17) (2020) 173001.
\newblock \href {http://arxiv.org/abs/1910.00588} {\path{arXiv:1910.00588}},
  \href {https://doi.org/10.1088/1751-8121/ab5d50}
  {\path{doi:10.1088/1751-8121/ab5d50}}.

\bibitem{Drukker:2010nc}
N.~Drukker, M.~Marino, P.~Putrov, {From weak to strong coupling in ABJM
  theory}, Commun. Math. Phys. 306 (2011) 511--563.
\newblock \href {http://arxiv.org/abs/1007.3837} {\path{arXiv:1007.3837}},
  \href {https://doi.org/10.1007/s00220-011-1253-6}
  {\path{doi:10.1007/s00220-011-1253-6}}.

\bibitem{Drukker:2008zx}
N.~Drukker, J.~Plefka, D.~Young, {Wilson loops in 3-dimensional N=6
  supersymmetric Chern-Simons Theory and their string theory duals}, JHEP 11
  (2008) 019.
\newblock \href {http://arxiv.org/abs/0809.2787} {\path{arXiv:0809.2787}},
  \href {https://doi.org/10.1088/1126-6708/2008/11/019}
  {\path{doi:10.1088/1126-6708/2008/11/019}}.

\end{thebibliography}
\end{document}